# Software engineering as a domain to formalize

Bertrand Meyer
Eiffel Software
October 2024 (revised February 2025)

*Abstract*—Software engineering concepts and processes are worthy of formal study; and yet we seldom formalize them. This "research ideas" article explores what a theory of software engineering could and should look like.

Software engineering research has developed formal techniques of specification and verification as an application of mathematics to specify and verify systems addressing needs of various application domains. These domains usually do not include the domain of software engineering itself. It is, however, a rich domain with many processes and properties that cry for formalization and potential verification.

This article outlines the structure of a possible theory of software engineering in the form of an object-oriented model, isolating abstractions corresponding to fundamental software concepts of project, milestone, code module, test and other staples of our field, and their mutual relationships. While the presentation is only a sketch of the full theory, it provides a set of guidelines for how a comprehensive and practical Theory of Software Engineering should (through an open-source community effort) be developed.

*Keywords—formal methods, theory of software, formal modeling, object-oriented modeling.* **This paper benefitted from help by Jocelin Fiat and Javier Velill**

Can software engineering apply one of the most powerful ideas it has developed over several decades, *formal specification*, to itself? This note suggests that formalization of software engineering concepts is a worthwhile endeavor and proposes basic elements of such a project – essentially, an ontology of the core concepts of our field – with the intent of starting, if there is enough interest, a community-based effort to produce a widely-accepted model and supporting tools.

Formal methods are mathematical techniques for specifying elements of a certain IT-related problem domain and serve as a basis for verifying that they possess certain properties. It is necessary for this definition to indicate "IT-related", since otherwise it would just characterize the scientific and engineering method in general (physics, for example, could then be recast as "formal methods" for studying certain natural phenomena).

The principal IT problem domain to which formal methods have been applied so far is programming: we specify the requirements of a system, meaning the objects that it manipulates and their desired properties, as a basis for verifying rigorously whether a proposed implementation (a program) satisfies those requirements.

Software engineering includes much more than programming, but little in it beyond programming has been formalized. (There have been a few notable attempts in this direction, some of which are cited in Section II.) The field could benefit from such systematic efforts to understand and describe the concepts of our field in a precise form permitted by formal methods. Any such formalization should include not only definitions of the basic concepts (such as product, team member, delivery, module, deadline, requirement element, test case, test suite, test oracle, milestone...) and axioms expressing fundamental properties (for example, a product on which a developer works must be part of a planned delivery).

This combination of definitions, axioms and theorems makes up what in science is called a *theory*, providing a precise formalization of a problem domain. We are used to theories in programming, with examples such as axiomatic (Floyd-Hoare-Dijkstra) semantics, model checking or abstract interpretation, as well as theories of algorithmic complexities. Here the aim is the same, but the objects of discourse are the basic constituents of the software engineer's job.

Modeling these "objects" will, in the developments below, rely on object-oriented techniques, whose basic guidance is to look for the right abstractions, here abstractions of software engineering, classifying these abstractions through inheritance, and specifying their properties through types and logical properties (contracts).

The present article is a step towards such a theory; it describes how such a formalization of the software engineering domain could look like. It does not, of course, provide the theory, a goal which would require a series of detailed articles each formalizing an area of software engineering, or perhaps a textbook, a kind of formal version of SWEBOK, the Software Engineering Book of Knowledge [1]. It is intended to provide some basic elements and foster further discussion and elaboration of actual, detailed theories which would stand a good chance of wide adoption.

Benefits to be expected of undertaking such efforts potentially include:

- The ability to define accepted "best industry practices" in a precise way.

- The ability to determine precisely whether actual practices conform to them, and to write tools that will perform such verifications.

- The ability to define both standard processes (say, waterfall or scrum) and organizations' own variants (with, for example, a company has defined its own process which is based on Scrum but includes elements of DevOps and company-specific extensions).

- The ability to determine whether the practice of a project actually meets the process specification, again with tools to support that verification.

- The ability to prove that certain systems or processes satisfy stated properties.

- Support for certification (for example, CMMI or ISO).

- The ability to develop better tools, in particular project management tools, as they can (unlike general-purpose tools supporting management of projects of any kind) rely on a precise model of software-specific concepts.

- Support for teaching software engineering in a more systematic and productive way.

- On a purely intellectual level, a better understanding of software engineering (as always follows, in any problem domain, from a formalization effort leading to a high-quality result).

PREVIOUS EFFORTS

The observation that software is worth formalizing is by itself not new; it goes back at least to Osterweil's classic article [2][3] from 1987, whose title is by itself a manifesto: "*Software processes are software too*". Osterweil did not attempt to formalize software in a mathematical way, but emphasized that software processes are worthy of systematic analysis. Interestingly, he found it necessary to design a language (in today's terminology, we would say a DSL, a Domain-Specific Language). The underlying theory is not, however, spelled out, and the fame of the paper has not led to the spread of tools relying on its concepts.

Another early attempt was the "Software Knowledge Base" [4], which sketched a relational theory of connections between software elements. It mostly focused, however, on modules and other program elements, rather than general artifacts and processes of software engineering.

SWEBOK, already cited, is a major achievement having codified much of the known understanding of software engineering and its best practices. SWEBOK is, however, largely informal. Typical of countless examples is the definition of "architecture evaluation":

> *3.2.3. Architecture Evaluation*
>
> Architecture evaluation validates whether the chosen solutions satisfy ASRs and when and where rework is needed. Architecture evaluation methods are discussed in topic 4 *Software Architecture Evaluation*.

Another difference with the goal of providing a theory of software engineering is that SWEBOK is not just descriptive but normative: it intersperses descriptions of software concepts with prescriptions of how to handle them according to industry best practices. That feature is part of the charter of SWEBOK but a theory must focus on the descriptive (not mixing "news" and "editorial"). Defining speed, as the quotient of distance traveled to time to travel it, comes separately from (and before) enacting speed limits.

The CMMI standard originating with the US DoD [5] also includes an extensive definition of principles and "disciplines", which provide a rich set of definitions of essential concepts of software engineering. Like SWEBOK, however, it remains at the level of English descriptions and does not come close to a full-fledged theory of the field.

The SEMAT effort [6] was proudly announced in 2009 in an article [7] proclaiming that "methods need theory" and that the field of software engineering requires a strong theoretical basis. The result so far has been the "Essence" methodology whose specification [8] includes useful definitions of basic concepts, for example (section 4 of that document):

> *A Method is the composition of a Kernel and a set of Practices to fulfill a specific purpose.*

The description, however, remains at the level of an English text, with no attempt at a more systematic formalization. In addition, Essence is not a full-fledged theory of software engineering but a method (somewhat paradoxically, since the original SEMAT manifesto [7] announced an attempt to end the proliferation of methods). Being a method rather than a theory, Essence is like SWEBOK prescriptive and not just descriptive. In addition it does not just address well-known concepts of software engineering but contains "original research" in the sense of Wikipedia [9] (which prohibits such elements in its own articles), such as "Alphas" ("Abstract-Level Progress Health Attributes" [6]), a powerful concept but not one that has yet gained wide acceptance in the field. We may expect of a theory that it will focus on classifying and specifying generally recognized fundamental notions of the problem domain.

All these efforts provide important definitions and analyses, which any attempt to formalize the field must take into account. They do not, however, provide the formalization itself.

CONVENTIONS

The rest of this presentation, while also not providing full formalization, will give some elements of what such a formalization will look like. Rather than fully formal, it combines elements of three kinds: explanations in plain English, graphical illustrations, and precise specifications. This approach, intended for readability, is inspired by an earlier article on "multirequirements" [10], which presents a specification methodology integrating these three levels of presentation.

One of the goals is to come up, after community discussion, to a widely accepted model – an ontology and taxonomy – of all the fundamental concepts and tools of software engineering, their properties and their mutual relations. The examples below are drawn from a first version which I developed with Jocelyn Fiat and Javier Velilla from Eiffel Software. It is in a GitHub repository that we are preparing to make public.

The precise specification part does not use a mathematical specification language, but a programming language also intended as a specification vehicle: Eiffel [11]. The advantage is to have a readable notation and to benefit from the structuring mechanisms of object-oriented modeling with classes, client-supplier relations between them, inheritance (for classification) and, to describe logical constraints, "contracts" (preconditions, postconditions, class invariants). The use of Eiffel as a specification formalism has been widely described (see e.g. [14]). The description could be expressed in another contract-equipped object-oriented notation such as JML [12] or, losing the object-oriented modeling facilities, a formal specification language such as Z [13].

The graphical notation is BON, Waldén's and Nerson's Business Object Notation [15] [16], a graphical notation for expressing object-oriented system structures. They can readily be translated into UML, but BON rests only on a small number of graphical conventions and supports Design by Contract concepts. The BON diagrams appearing below are produced automatically from Eiffel code (or the other way around) by the EiffelStudio IDE [17].

The style of the presentation can be described as "mock tutorial": we give a few glimpses of what a theoretical presentation of selected software engineering concepts would look like. As mentioned, the presentation is very partial (hence "glimpses"); in addition, while we expect the reader of the present article to know (for example) what a software project and a milestone are, the idea is to sketch how a full-fledged theoretical description would present the entire field to a student discovering it (the goal of any comprehensive theory).

The various levels (English, graphical, formal) can refer to each other. In the English text, the phrase "an A" where A is the name of a class in the formal text means "an instance of A", for example "a MODULE" or "a PROJECT". (We do not change in the plural, e.g. "two MODULE" without an "s".)

**PROJECTS**

Although it is possible to enter the software engineering world from many sides, one of the universals of the field is the notion of project. Fig. 1(after the bibliography) shows the overall "project" cluster. (A cluster is a group of classes, also called a package in some OO languages.) Here are some multirequirements-style elements of explanation.

A project is intended to produce a certain collection (SET) of PRODUCT and has a sequence (LIST) of MILESTONE:

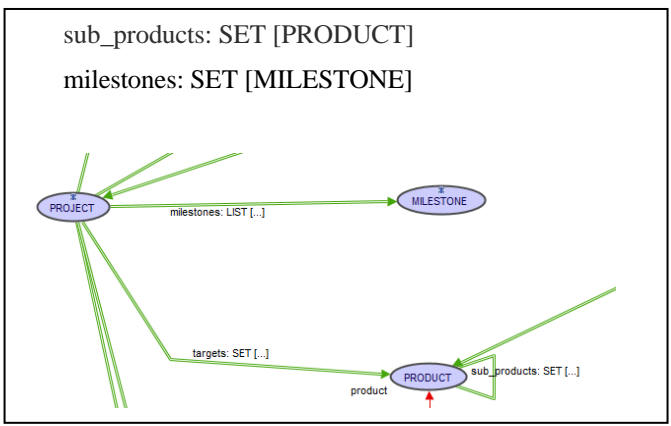

A PRODUCT, as shown, can have sub-products:

sub_products: SET [PRODUCT]

A project MILESTONE is defined by a set of PRODUCT_INCREMENT that the milestone must produce. A PROJECT_INCREMENT (which could also be called PRODUCT_VERSION) can be a new product, a PRODUCT_CREATION, appearing with the particular milestone, or a PRODUCT_UPDATE providing a new version of a product that was already present.

```
class PRODUCT_UPDATE
    inherit PRODUCT_INCREMENT feature ... end
class PRODUCT_CREATION
    inherit PRODUCT_INCREMENT feature ... end
```

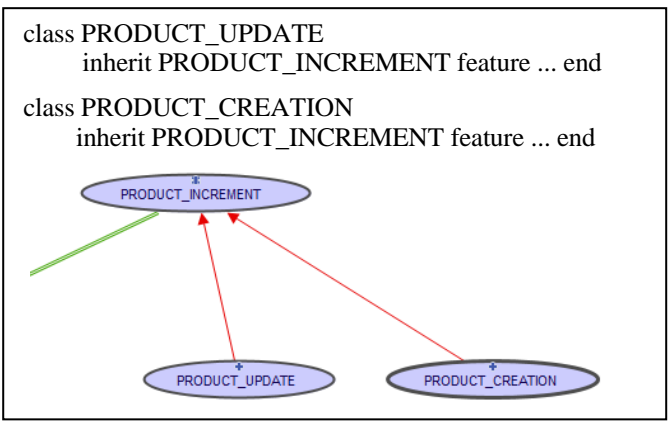

An important benefit of the OO approach to modeling relying on inheritance is that it does not require us to list possibilities exhaustively, as in "a product increment is *either* a product update or a product creation". Producing such closed lists is dangerous in the process of building a theory, as it makes it hard to add new variants later on: each time there is the risk of having to revisit and update many other parts of the theory that relied on the knowledge of the exact initial list. With inheritance (applying the "Open-Closed Principle"), we do not close such sets of variants of a basic notion but list variants individually: a PRODUCT_UPDATE is a kind of PRODUCT_INCREMENT; a PRODUCT_CREATION is another kind; and so on, but we do not preclude adding new kinds later on as the theory develops and covers more ground. Such additions do not normally imply updating the parts of the theory that have already been developed.

A PRODUCT_INCREMENT of any kind only makes sense if it is relative to a PRODUCT addressed by the project. (The PRODUCT is a software artifact, such as a code MODULE or a TEST_PLAN; a PRODUCT_INCREMENT is a version of that PRODUCT). One of the roles of producing a theory of software engineering is to specify such important properties formally. With object-oriented modeling, they can be expressed as clauses of class invariants; for example, in class PRODUCT_INCREMENT:

$$\exists\, e: current\_project.milestones \mid$$
$$product.milestone \in e.elements$$

In words: at least one of the PRODUCT that have been specified among the targets of the current MILESTONE must be the PRODUCT for which the current object is a PRODUCT_INCREMENT.

A key part of the theory will consist of specifying such fundamental consistency constraints, which, together with the definition of basic software engineering types (PRODUCT etc.) make up the basic competence of a professional software engineer.

**CLUSTERS**

The "project" cluster sketched above is one of the basic clusters of the current model. Each cluster covers an important part of the field. Current ones include: Project, Plans, Documents, Bugs, Events, Issues, Messages, People (TEAM, TEAM_MEMBER, STAKEHOLDER etc.), PRODUCTS (with subclusters including Code and Tests), Tasks, Support, Discipline (in the CMMI sense) and Processes.

**MODELING PROCESSES AND BEST PRACTICES**

The classes of the Processes cluster model the notion of PROCESS (one of its basic classes) and associated abstractions. Here we only indicate how the present theoretical framework handles these notions.

The natural inclination is to model a PROCESS (for example Waterfall, Spiral, RUP, Scrum…) through its components: REQUIREMENTS_PHASE, DESIGN_PHASE etc. for the Waterfall; CYCLE, PROTOTYPE etc; SPRINT, DAILY_MEETING, SPRINT_RETROSPECTIVE etc. (plus artifacts such as BURNDOWN_CHART) for Scrum.

Closer analysis suggests, however, that this approach is *not* the best way to handle the notion of software engineering process. The abstractions in question, such as the examples cited, are all important but they do not define a process. They are important software engineering abstractions with definitions of their own, independently of how a particular process variant combines them. Another fundamental property of the "process" abstraction is that a process defines how the organization *wants* to conduct its software business, but it is an inevitable fact of life that what the organization actually *does* will not always match what it wants to do. In OO modeling terms, a fundamental feature of the PROCESS abstraction (yielding a method of the corresponding class and its descendants) is that a process, or some element thereof, can be followed or not.

These observations suggest that the theory should treat a PROCESS as a constraint, defined by a boolean-valued method specifying (if it has value True) that elements of the project have been conducted in a certain way. Class PROCESS_RULE, the top of an inheritance hierarchy with many variants describing process elements, introduces an abstract function *constraint* which defines such rules. An example constraint for the waterfall is

```
design_phase.start_time ≥ requirements_phase.end_time
```

This property is *not* a class invariant (an axiom of the theory), which would express that all projects everywhere observe it! Even if it only expresses that some projects, or just one specific project will, that approach is not realistic, as it describes hopes (wishful thinking) rather than a guaranteed reality. As noted in section II, a theory of a domain of interest should be descriptive before it becomes normative. More precisely, its approach to normative rules should be descriptive too: the theory specifies the rules and mechanisms to determine whether other objects of the theory (for example, in the software case, project phases) observe them or not.

In the preceding example, a specific class WATERFALL_PROCESS (inheriting from PROCESS) will specify a constraint given by the boolean expression above. PROCESS and its descendant classes have a function *is_satisfied* (… *arguments* …) which assesses whether components of the process (given by the arguments) satisfy the constraint.

Process modeling is one of the areas where the power of OO modeling pays off. One of the features of process models – other than the property, just noted, that in real life as opposed to textbooks no project *ever* follows any process model *exactly*, as is to be expected of human-driven phenomena – is that no company ever adopts a recommended process model exactly: each organization adapts the model, be it Waterfall, Spiral, RUP, XP, Lean, Scrum or any other, to its own needs, constraints and company culture. The object-oriented model (through its "Open-Closed Principle") makes it possible to define new classes, say OUR_SCRUM_VARIANT, which inherit from a predefined one such as SCRUM_PROCESS and, using the full power of inheritance, keep what remains applicable and redefine ("override") what needs to be adapted.

One of the goals of the present project is to come up, as part of the theory, with a *library* of classes covering the major, best-known process models (as listed above), open to individual adaptation, through inheritance, by organizations having defined their own process specifics.

The general idea of specifying processes as constraints applies more broadly to the description of principles, practices and disciplines of software engineering

VERIFICATION

Another important application of a theory is to enable systematic verification of candidate solutions against a specification. In the case of programs, while testing remains the usual form of (partial) verification in many industry circles, more static techniques are also gaining ground. Theories of programming, such as axiomatic semantics, provide the basis for verification toolsets, nowadays quite sophisticated, to verify the correctness of programs. An example among many is Boogie [18]; others include the numerous existing tools for model checking and abstract interpretation.

In the same way, the axioms and theorems of a Theory of Software Engineering are subject to verification. As with programs, one can use testing (executing simulation runs of processes and monitoring preconditions, postconditions and class invariants) or, more systematically, static verification.

In this respect it is important to note that the object-oriented techniques used above are compatible with formal verification. This article started with goals of formal verification and proceeded to define an object model, but there is no contradiction between the two: modern object-oriented languages supporting specification (Design by Contract techniques), such as Eiffel, JML and Dafny [19] are just as formal as classical notations officially recognized as "formal specification languages"; as a result, they come equipped with a verification (proof) infrastructure. The development of the AutoProof framework based on Boogie [20] applies this idea (in addition to the dynamic tests supported by other tools, based on monitoring contracts at run time) to the static verification of properties of theory and of individual processes.

It is also important to point out that many formal properties do not require a logical expression (a boolean expression in a precondition, postcondition or class invariant) but can simply be expressed – as examples have shown – through type properties. We do not need for example a formal specification of the property "the targets of a project are products": we simply declare targets, in class PROJECT, as being of type SET [PRODUCT]. Numerous properties are implicitly specified this way, taking advantage of the compiler's type checks (for a powerful type system with generics and inheritance) as verification. True proof logical specifications and the associated sophisticated verification mechanisms of a program prover are reserved for advanced logical constraints.

FUTURE WORK AND CONCLUSION

As noted earlier, we (Jocelyn Fiat, Javier Velilla and I) have developed a first version of the object model (the ontology), from which the above extracts are taken. The model consists of a set of classes, describing abstractions rather than implementations; the diagram extracts shown above are part of the overall diagram produced automatically by EiffelStudio.

For the moment the repository is private, mostly because we do not know whether anyone else is interested. (Also because it needs a bit of cleanup before it goes ballistic.) If interest there is, we will make the repository public; I will provide the information in a future post.

## FUTURE WORK AND CONCLUSION

The representative but partial extracts shown in previous sections are part of an ongoing work to cast the fundamentals of software engineering into a systematic theory in the form of an object model. A first version is available for the basic clusters (Project, Code, Issues, Tasks, Events, Bugs). Work is continuing on the rest. It follows the principles stated earlier: description rather than prescription (and prescription itself, that is to say, normative elements, covered descriptively too); "no original research" unless strictly necessary (in other words, we are not out to impose yet another methodology on the software engineering world, but to describe basic concepts and allow originators or proponents of any methodology to describe it precisely); incremental development taking advantage of reuse (libraries) and inheritance; use of a sophisticated OO type system; use of formal logical properties enabling verification by a program prover; and, throughout, focus on isolating and describing the key abstractions defining the field of software engineering.

We have started the effort and brought it to a first level of presentation and verification, but no single team has the breadth of software engineering competence that would make it possible for the theory to cover the field. The effort is open-source and explicitly includes for all interested members of the community to bring their expertise. Success would mean that we can at last achieve the goal, often stated but never realized, of treating the engineering software into a well-defined domain, worthy of scientific study and covered by a useful theory.

Acknowledgments In addition to the work of Jocelyn Fiat and Javier Velilla, this article benefitted from thorough comments by Jean-Michel Bruel.

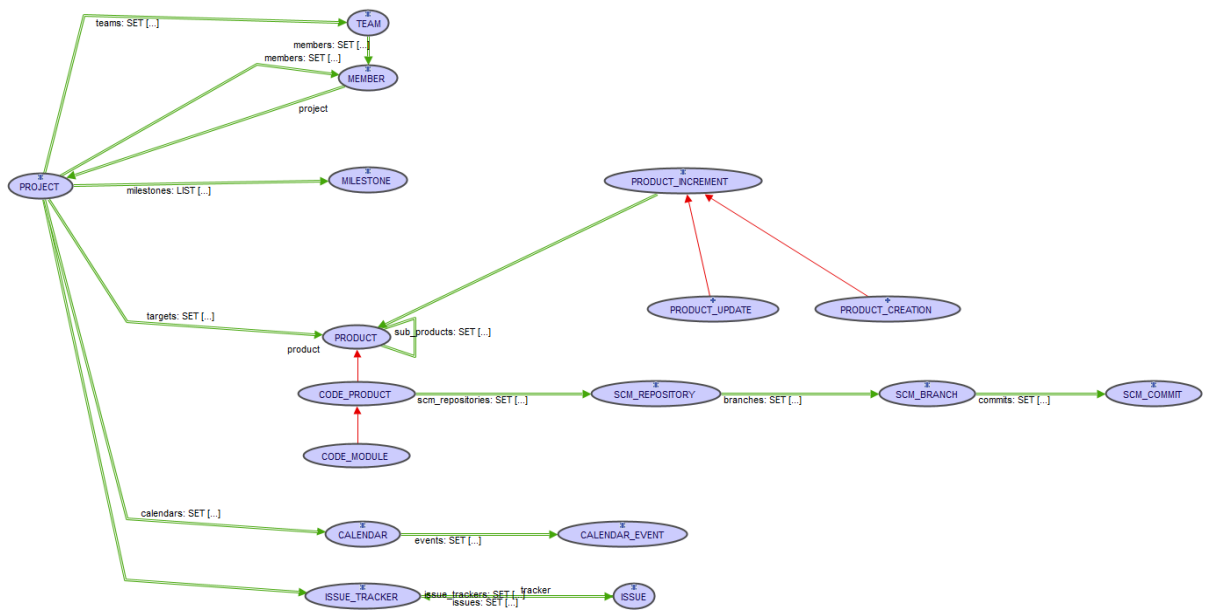

Fig. 1. The PROJECT cluster, overall structure